# *ROSAT* PSPC observations of Cygnus-A : X-ray spectra of the cooling flow and hot spots.

C. S. Reynolds and A. C. Fabian
*Institute of Astronomy, Madingley Road, Cambridge CB3 0HA*

**ABSTRACT**
We present a *ROSAT* Position Sensitive Proportional Counter (PSPC) observation of the powerful radio galaxy Cygnus-A. The X-ray emission in the *ROSAT* band is dominated by thermal emission from the hot intracluster medium of the associated cluster. Image deprojection confirms the existence of a significant cluster cooling flow with total mass deposition rate of $\sim 250\,{\rm M_\odot\,yr^{-1}}$ and a (Hubble time) cooling radius of $\sim 180\,{\rm kpc}$. Spectral data show the associated gradient in the emission-weighted mean temperature with the temperature decreasing towards the centre of the cluster. We also find signatures of the radio source: in particular, we detect the X-ray emission from the western radio hot spot previously found by the *ROSAT* High Resolution Imager (HRI). We find the emission from the hot spot to be hard and discuss the physical implications of this result.

## 1 INTRODUCTION

The powerful radio galaxy Cygnus-A (3C 405) is often considered to be the archetypal classical double. It is the brightest extragalactic radio source in the sky and displays a pair of spectacular radio lobes (Hargrave & Ryle 1974; Alexander, Brown & Scott 1984; Carilli et al. 1991). These lobes are thought to be powered by an obscured quasar (Djorgovski et al. 1991; Ward et al. 1991) via highly collimated supersonic/relativistic jets. Optically, Cygnus-A is identified with a cD galaxy (z=0.0574) in a supposedly poor cluster of galaxies (Spinrad & Stauffer 1982).

X-ray emission from Cygnus-A was first detected by the *Uhuru* satellite (Giacconi et al. 1972). Since then it has been observed by every major X-ray mission. *Einstein Observatory* High Resolution Imager (HRI) observations (Arnaud et al. 1984) resolved the emission from the intracluster medium (ICM) and revealed the presence of a cooling flow with a mass deposition rate of $\sim 90\,{\rm M_\odot}\,yr^{-1}$. Observations with the *EXOSAT* medium-energy telescope (ME) strongly suggested the presence of a power law component in the 1–20 keV spectrum (Arnaud et al. 1987). *Ginga* spectra confirm the presence of this power law component and require it to be highly absorbed (Ueno et al. 1994). The power law photon index ($\Gamma = 1.98 \pm 0.2$) and the absorbing column density ($N_{\rm H} = 3.7 \pm 0.7 \times 10^{23}\,{\rm cm^{-2}}$) are consistent with this component originating from an obscured (type II) quasar. Most recently, *ROSAT* HRI observations show clear evidence for hydrodynamic interaction of the jets/lobes with the ICM as well as emission from the radio hot spots themselves (Harris, Carilli & Perley 1994; Carilli, Perley & Harris 1994).

Such observations show that the X-ray emission from Cygnus-A is complex with components possibly originating from the active galactic nucleus, hydrodynamic interaction of the jets with the ICM and the radio hot spots (via synchrotron self Compton emission; SSC) as well as the thermal emission from the ICM and the cooling flow. To disentangle these components requires spectral information as well as spatial information, such as that provided by the *ROSAT* PSPC.

This paper presents a *ROSAT* PSPC observation of Cygnus-A. In particular, we address the properties of the ICM. Section 2 briefly describes the observation. Section 3 presents spatial analyses of the data including a deprojection analysis and a hardness-ratio study. A deprojection of the *ROSAT* HRI data is also presented. In addition to thermal emission from the ICM and confirmation of the cooling flow, we find hard emission from the vicinity of the western hot spot. Section 4 investigates the spectral properties the various emissions. Physical models for the emission from the western hot spot (given the constraints imposed by this observation) are discussed in Section 5. Section 6 provides a summary of the issues raised.

## 2 OBSERVATIONS

The *ROSAT* PSPC has a bandpass of $\sim 0.1$–2.5 keV over a 2 degree diameter field of view. It has a spatial res-olution of ∼25 arcsec (FWHM) and an energy resolution of $\Delta E/E = 0.43(E/0.93)^{-0.5}$ (FWHM). The PSPC background count rate is very low and mainly dominated by cosmic ray induced particle background and scattered solar X-rays. The particle background is efficiently rejected leaving a residue count rate of $\sim 5 \times 10^{-6}$ counts s$^{-1}$ arcsec$^{-2}$ keV$^{-1}$ (Snowden et al. 1992).

Cygnus-A was observed by the *ROSAT* PSPC on 1993 October 10/11. The total (on-source) exposure time for this observation is $\sim 9\,400$ s. The observations were performed in the standard 'wobble' mode (with ∼400 s period) to avoid shadowing of sources by the coarse wire grid which covers the PSPC aperture. Periods of high particle background were rejected, leaving just over 9 000 s of usable data.

Soft X-ray observations of Cygnus-A are severely affected by its low Galactic latitude which results in a high column density, $N_{\rm H}$, of cold Galactic material along our line of sight to this source. Studies based on optical extinction conclude a column of $N_{\rm H} \approx 2.4 \times 10^{21}$ cm$^{-2}$ (Ueno et al. 1994 and references therein). However, we assume a Galactic column of $N_{\rm H} = 3.2 \times 10^{21}$ cm$^{-2}$. This higher value is suggested by measurements of the 21-cm line of H I at nearby positions on the sky (Stark et al. 1992) and is more consistent with our spectral analysis (Section 4).

Throughout this work we assume that $H_0 = 50$ km s$^{-1}$ Mpc$^{-1}$ giving a distance to Cygnus-A of 340 Mpc. At this distance, 1 arcmin corresponds to 100 kpc.

Initial data reduction was performed using the FTOOLS software package.

## 3  IMAGE ANALYSIS

Here we investigate the spatial distribution of X-ray emission.

### 3.1  Full band images

There are very few source counts below 0.5 keV. Thus, to reduce background in our image analyses, the data below 0.5 keV were ignored. Images were extracted in the 0.5–2.5 keV band and finely binned into 4 arcsec pixels. An adaptive smoothing algorithm (Rangarajan et al. 1995) was then applied so that the local smoothing length contained at least 25 counts. Structure in the resultant images is then (formally) significant at the 5$\sigma$ level.

Fig. 1 shows the full band image of Cygnus-A. The region shown is approximately that of the inner ring of the PSPC. Thermal cluster emission dominates this image and is seen to be strongly peaked on the position of Cygnus-A. On large scales there are gross deviations from spherical symmetry. An extensive 'plume' is seen to reach at least 15 arcmins (1.5 Mpc) north-west of Cygnus-A. Isophotes in the central region also appear distorted along a position angle coincident with that of the radio axis. Fig. 2 shows the X-ray emission in the central regions of the cluster overlaid on a 4.5 GHz Very Large Array (VLA) map of Cygnus-A. The distortion of the isophotes within the central 1 arcmin radius (i.e. the extent of the radio source) is readily interpreted as emission from the hot spots and hydrodynamic interaction of the radio emitting plasma with the ICM, as discovered by the *ROSAT* HRI (Harris, Carilli & Perley 1994; Carilli, Perley & Harris 1994).

**Figure 1.** *ROSAT* PSPC image of the Cygnus-A region in the 0.5–2.5 keV band. This image has been obtained by finely binning the data into 4 arcsec pixels and then adaptively smoothing so that the local smoothing length contains at least 25 counts. Thus, features in this image are (formally) significant at the 5$\sigma$ level. Note the plume extending over 1.5 Mpc to the north-west of Cygnus-A itself.

**Figure 2.** 4.5 GHz VLA map of Cygnus-A (greyscale) with logarithmically spaced contours of total X-ray intensity (from *ROSAT* PSPC). The X-ray data has been adaptively smoothed to the 5$\sigma$ level, as in Fig. 1. Note the alignment between the radio source and the asymmetries in the X-ray morphology. This is readily interpreted in terms of X-ray emission from the radio hot spots and hydrodynamic interaction of the radio source with the ICM (Harris, Carilli & Perley 1994; Carilli, Perley & Harris 1994).

### 3.2  Deprojection analysis and the cooling flow properties

The technique of image deprojection has been highly productive in the study of the ICM of numerous clusters (Fabian et al. 1981; Arnaud 1986; White 1992; Allen et al. 1993;

**Figure 3.** Deprojection results for *ROSAT* PSPC and *ROSAT* HRI data of the Cygnus-A region. The PSPC data extends out to a radius of 600 kpc whereas the HRI data is only used out to 80 kpc (due to the higher background in the HRI). The assumptions and input parameters required for this deprojection are discussed in the main text. A cooling flow with total mass deposition rate $\sim 250\,{\rm M}_\odot\,{\rm yr}^{-1}$ is confirmed by this data. The spatial region over which the mass deposition occurs is well resolved by the HRI data.

White et al. 1994). Assuming spherical symmetry, quasi-hydrostatic equilibrium and a form for the gravitational potential of the cluster many properties of the ICM can be obtained including the run of density, temperature and mass deposition rate with radius. As demonstrated above, the ICM of Cygnus-A has clear deviations from spherical symmetry. However, the *underlying* ICM emission (i.e. excluding the effects of the interaction with the radio emitting plasma or lobes) in the central 500 kpc of this cluster appears approximately circular. Assuming this translates into a spherical distribution of emitting gas, this region of the cluster is open to the deprojection method.

A radial X-ray surface brightness profile was made along a position angle perpendicular to the radio axis in an attempt to approximate the underlying ICM emission. We assume a gravitational potential comprising of two King models, one of which represents the cluster potential and the other which represents the potential of the cD galaxy. For the cluster potential, we use a line-of-sight velocity dispersion of $900\,{\rm km\,s}^{-1}$ and a core radius of 150 kpc. Following Arnaud et al. (1984), we take the low end of the measured optical velocity dispersion (Spinrad & Stauffer 1982) in order to obtain consistency between the deprojected data and the cluster temperature of 7.3 keV as measured by *Ginga* (Ueno et al. 1994). The parameters of the potential of the cD galaxy do not sensitively affect the deprojection since any reasonable galaxy core radius would lie well below the spatial resolution of *ROSAT*. Thus, we take the typical values of 500 pc and $300\,{\rm km\,s}^{-1}$ for the core radius and velocity dispersion respectively (Arnaud et al. 1984).

The above deprojection analysis was performed on both *ROSAT* PSPC and *ROSAT* HRI data. PSPC data are used to a radial distance of 600 kpc from the centre of the cluster whereas HRI data are only used to 80 kpc (due to the higher background in the HRI). During the deprojection of the HRI data, the pressure was fixed to agree with the PSPC deprojection at the 80 kpc point. The results of the deprojections are shown in Fig. 3. The two instruments are in excellent agreement and the deprojection confirms the presence of a significant cooling flow with a Hubble time cooling radius (i.e. the radius at which the cooling time equals the Hubble time) of $\sim 180$ kpc and a total mass deposition rate of $\sim 250\,{\rm M}_\odot\,{\rm yr}^{-1}$. The difference between our result and that of Arnaud et al. (1984) is mainly due to the difference in the assumed absorbing Galactic column density. Our result is in good agreement with a recent deprojection of the *Einstein* observatory image (White 1992).

The HRI data resolve well the spatial region over which the mass deposition occurs. Most of the mass deposition occurs in the inner 50 kpc of the cluster for which the cooling time is less than $3 \times 10^9$ yr. This could be indicative of the time since the cooling flow last suffered a major disruption event (possibly due to a cluster-cluster merger). The brightess profile follows the expected $r^{-1}$ behaviour within this radius and $r^{-2}$ outside this radius. We note the similarity with the Perseus cluster (Edge, Stewart & Fabian 1992).

Cygnus-A is the most dramatic example of a low-redshift high-power AGN in the centre of a cooling flow. It is pertinent to discuss whether the AGN activity could place enough energy into the ICM to (on average) offset the bolometric cooling luminosity, $L_{\rm cool}$. For many clusters (including the Cygnus-A cluster) $L_{\rm cool}$ can approach, or exceed, $10^{45}$ erg s$^{-1}$. Thus, over a Hubble time the ICM emits thermal radiation in excess of $3 \times 10^{62}$ erg. For AGN activity to offset this energy loss (and so be energetically important for the ICM), *all* of the accretion energy of a $10^9$ M$_\odot$ black hole would have to be transferred into the ICM. In practice, only a small fraction of the total AGN power would be taken up by the ICM (for example, via shocks at the head of expanding radio lobes) and the resulting black hole masses required for the AGN activity to dominate the ICM would become implausibly large. We note that the *instantaneous* kinetic luminosity of the jets could exceed the cooling luminosity without violating any constraints based on the black hole mass. The subsequent affect on the ICM would depend on the magnitude of any AGN outburst and how effectively the energy could be distributed and dissipated into the ICM.

### 3.3  Hardness-ratio study

The *ROSAT* PSPC gives imaging and spectral information simultaneously. Thus we can examine the hardness of the emission across the image.

Images were extracted in the 0.5–1.3 keV and 1.3–2.5 keV bands (using 4 arcsec pixels). These bands were chosen so as to achieve roughly equal count rates in each image. The two images were adaptively smoothed to the $5\sigma$ level and are displayed in Fig. 4. The most striking difference between the two bands is the western emission detected in the hard band. This emission is positionally coincident with the western radio hot spot and is identified with the hot-spot emission found by the *ROSAT* HRI (Harris, Carilli & Perley 1994). Thus we find the western hot spot X-ray emission to be significantly harder than the cluster emission. Constraints on its spectrum are discussed in Section 4.2. We fail to detect clearly a corresponding feature coincident with the eastern hot spot. We attribute this to the limited spatial resolution of the PSPC which leads to emission from the eastern hot spot being inseparable from the strongly peaked thermal ICM emission: the eastern hot spot is $\sim 15$ per cent closer to the centre of Cygnus-A, where the surface brightness of the cluster emission is 50 per cent higher.

| Region | Temperature (keV) | $\chi^2$/dof |
|---|---|---|
| 5 arcmin radius | $3.2 \pm 0.6$ | 54/59 |
| 1 arcmin radius | $2.5^{+0.7}_{-0.4}$ | 30/35 |
| 1–2 arcmin annulus | $3.7^{+2.8}_{-1.2}$ | 16/30 |
| 2–3 arcmin annulus | $3.4^{+4.1}_{-1.4}$ | 15/18 |
| 3–4 arcmin annulus | $> 3.3$ | 14/19 |
| 4–5 arcmin annulus | $> 4.0$ | 25/32 |
| NW plume | $> 4.1$ | 22/43 |

**Table 1.** Results of spectral fitting to regions of the Cygnus-A cluster. For each region, a one component Raymond-Smith plasma model modified by Galactic absorption has been fitted. The Galactic column density was fixed at $N_{\rm H} = 3.2 \times 10^{21}$ cm$^{-2}$ and the plasma abundance was fixed at $0.3$ Z$_\odot$. Errors are quoted at the 90 per cent confidence level for one interesting parameter ($\Delta\chi^2 = 2.7$).

## 4  SPECTRAL ANALYSIS

Detailed spectral information complements spatial information when attempting to disentangle a complex system such as Cygnus-A. This section presents a spectral analysis of Cygnus-A. Galactic absorption severely affects the observed soft X-ray flux from this object. Leaving the Galactic column density of cold absorbing matter, $N_{\rm H}$, to be a free parameter in the spectral fits leads to a very degenerate $\chi^2$ space and very poor constraints on parameters such as the temperature of the ICM. Thus we choose to fix $N_{\rm H}$ (thereby assuming uniform absorption over the whole PSPC field of view). Spectra are background subtracted using a background field from the same PSPC observation.

Unless otherwise stated, all errors and limits are quoted at the 90 per cent confidence level for one interesting parameter, $\Delta\chi^2 = 2.7$.

### 4.1  ICM emission

Spectra were extracted in 1 arcmin annuli from the centre of Cygnus-A to a maximum radius of 5 arcmins (500 kpc). Spectra were also extracted for the whole inner 5 arcmins and the north-western plume (discussed Section 3.1). The Raymond-Smith model for emission from a thermal plasma (Raymond & Smith 1977) was fitted to these data (including the effect of Galactic absorption). The resulting best-fitting temperature is an emission-weighted mean of the temperature of the various thermal components present. Such fits do not constrain the abundance of the plasma due to the severe Galactic absorption at low energies. Thus the abundance is fixed at a typical value of $0.3$ Z$_\odot$.

Table 1 shows the results of our spectral fits. Fig. 5 shows the temperature as a function of radius for the selected annuli. A systematic increase in temperature with increasing radius is seen. Such behaviour is qualitatively expected within the cooling flow scenario where the central regions contain large quantities of material in a cooler phase as compared with the outer regions of the cluster (Allen et al. 1993). Spectra from the outer regions of the cluster, where the background contributes significantly to the total observed counts, only provide lower limits on the plasma temperature due to the lack of high energy response in the PSPC.

**Figure 4.** *ROSAT* PSPC images of the central regions of the Cygnus-A cluster in a) the 0.5–1.3 keV band and b) 1.3–2.5 keV band. These bands were chosen so as to approximately bisect the total count rate. Each image has been adaptively smoothed to the $5\sigma$ level. Note the western feature in the hard image which is very weak in the soft band. This feature is spatially coincident with the western hot spot.

**Figure 5.** Best fitting plasma temperature as a function of radius for annuli centred on Cygnus-A. The dashed line shows the mean *Ginga* temperature for the cluster (at $k_B T = 7.3$ keV). There is a clear gradient with the emission-weighted mean temperature decreasing towards the centre of the cluster. This is as expected from a cooling flow (see text).

The north-western plume appears significantly hotter than the more central regions of the cluster. However, if this emission represents the remnants of a merging cluster, we would expect it to be relatively cool on the basis of the known positive correlation between cluster luminosities and temperatures (Edge & Stewart 1991). Thus, either there is a significant departure from this correlation or some more complex physical process has led to the observed plume.

### 4.2 Western radio-lobe emission

In Section 3 we show evidence for hard X-ray emission from the vicinity of the western radio lobe. Here, we perform direct spectral studies of this emission.

A spectrum was extracted for counts in the immediate vicinity of the hot spot (with an extraction radius of ∼15 arcsec). Approximately 150 counts were collected from this region. The 'background' spectrum was extracted from a narrow annulus of the cluster emission at the same radius as the western hot spot but excluding the vicinity of both the western and eastern hot spots. The resulting background subtracted data represent the emission from the hot spot (without the associated cluster emission). The derived hot spot luminosity (in the 0.5–2 keV band) is $(9\pm 3)\times 10^{41}$ erg s$^{-1}$, in good agreement with Harris, Carilli & Perley (1994; hereafter HCP).

The statistical quality of the background subtracted spectrum is poor and thus only limits can be derived. First, a model consisting of emission from a single temperature thermal plasma (modified by Galactic absorption) was fitted to the data. The fit was statistically acceptable with $\chi^2 = 4.6$ for 5 dof. However, the temperature is poorly constrained ($T > 1.3$ keV at more than the 90 per cent confidence level). Secondly, a power law model with photon index $\Gamma$ (modified by Galactic absorption) was fitted to the data. The fit was excellent ($\chi^2 = 2.7$ for 5 dof) and gave the constraint $\Gamma < 1.7$ at the 90 per cent confidence level. These limits are insensitive to reasonable changes in the Galactic column density.

## 5 PHYSICAL MODELS FOR THE X-RAY EMISSION FROM THE HOT SPOTS

The observed luminosity and hardness of the X-ray emission from the vicinity of the radio lobes provide important constraints on any physical model for this emission. Here we examine possible physical models for this emission given the observational constraints of Section 4.2.

Thermal models for this emission are heavily constrained by radio polarization data (Dreher, Carilli & Perley 1987). HCP show that an explanation based on thermal emitting plasma within the radio lobe is inconsistent with the observed polarization (since the required amount of thermal material would result in Faraday depolarization of the radio emission). They also reject the hypothesis of thermal emission from a shocked sheath of material outside the radio lobe on the basis of the lack of increase of rotation measure (RM) over the hot spot as compared to the rest of the radio lobe. Such an increase in RM would be expected given this hypothesis due to the presence of magnetized, shocked gas along the line of sight to the radio hot spot. This assumes that the ICM possess a rather large intrinsic magnetic field, as suggested by Dreher, Carilli & Perley (1987). HCP conclude that the X-ray emission is SSC emission from the radio emitting plasma in the hot spots.

### 5.1 Non-thermal models

The SSC model of HCP assumes a cylinderical geometry for the hot spot with dimensions taken from the high resolution radio maps of Carilli, Dreher & Perley (1988). They also assume that the electron energy spectrum in the hot spots is a broken power law adjusted so as to produce the observed radio spectrum. For a given magnetic field (assumed uniform) the SSC spectrum is then determined. They can explain the observed X-ray flux with a magnetic field of $B \sim 150 \mu G$ (close to the equiparition value). In the *ROSAT* band, this model predicts a power law spectrum with photon index $\Gamma \sim 1.8$ (HCP). This is formally *inconsistent* with our spectral data at the 90 per cent level. Here we examine the physical implications of this result: in particular, we suggest a modification to the picture of HCP which can reconcile the SSC model with the current spectral constraint. We also discuss how other models fail (independently of the constraints imposed by RM studies).

Changing the geometry and/or size of the hot spot can have a significant affect on the observed emission. For example, suppose the observed hot spot was a disc seen in projection rather than a cylinder. The radio energy density within the hot spot would be less than assumed by HCP and thus a weaker $B$ field would be required in order that the SSC emission produced the observed X-rays. This has the effect of pushing the spectral break in the high-energy SSC spectrum to higher energies (see Fig. 3 of HCP). Placing the spectral break sufficiently close to the *ROSAT* band would resolve the discrepancy between the SSC model and the current spectral data.

There may be another source of photons which can be inverse Compton scattered to X-ray energies. The cosmic microwave background has an energy density an order of magnitude too low to produce the observed high energy emission unless the $B$ field in the hot spot is much lower than the equipartition value. However, if the hot spots experience relativistically beamed optical radiation from the central active nucleus, the optical photon energy density may be comparable to (or greater than) the radio energy density. Inverse Compton scattering of the optical photons to the *ROSAT* band would involve electrons well below the break in the electron energy spectrum. Assuming the electron energy spectrum derived from radio observations can be extrapolated to the lowest energy electrons, the energy density in these electrons is low leading to a small observable SSC flux. Thus, this would *not* be a plausible mechanism for the observed *ROSAT* band emission, although it might then be relevant to $\gamma$-ray emission from the hot spots resulting from inverse Compton scattering of the optical photons by electrons with energies near the break energy (i.e. the energy for which the energy density of the electron population is the greatest). Inverse Compton scattering of these optical photons could be relevant to emission in the *ROSAT* band if there exists an additional population of low energy electrons.

### 5.2 Thermal models

We now examine thermal models for the emission without imposing any constraints from RM studies (Dreher, Carilli & Perley 1987; Carilli, Perley & Dreher 1989). First, consider a smooth (one-phase) ICM. At the position of the hot spot, the above deprojection (Section 3.2) gives an ICM density of $n \approx 0.01 \, \text{cm}^{-3}$ and a temperature of $T \approx 5 \times 10^7$ K (corresponding to approximately 5 keV). We now examine the expected X-ray emission when this material is shocked by the advancing radio hot spot. Let the shock have Mach number $\mathcal{M}$ (with respect to the unshocked ICM). Assuming the lobes to be ram pressure confined gives $\mathcal{M} \approx 5$ (Carilli, Perley & Harris 1994). For a non-radiative, strong shock (i.e. $\mathcal{M} \gg 1$)

$$n_2 = 4n \tag{1}$$

$$T_2 = \frac{5}{16}\mathcal{M}^2 T \tag{2}$$

where $n$ and $T$ are the (number) density and temperature of the unshocked ICM and $n_2$ and $T_2$ are the (number) density and temperature of the shocked ICM. These relations assume a monatomic ideal gas.

Suppose the region of shocked, emitting material has volume $V$ which will be determined by the stand-off distance of the shock and the flow time of the material away from the region of high compression. We take the area of the shocked region to be $\sim 50 \, \text{kpc}^2$ (close to the upper limit implied by the inability of the *ROSAT* HRI to clearly resolve the emission) and the thickness of the shock to be $\sim 2 \, \text{kpc}$, giving $V = 100 \, \text{kpc}^3$. The observed thermal bremsstrahlung emission will have a isotropic luminosity between energies $E_1$ and $E_2$ of

$$L = \mathcal{L} V n_2^2 T_2^{1/2} \delta \tag{3}$$

where we define $\delta = e^{-E_1/kT_2} - e^{-E_2/kT_2}$ and $\mathcal{L} \approx 1.4 \times 10^{-27} \, \text{erg s}^{-1} \, \text{cm}^3 \, \text{K}^{-1/2}$ (for a simple hydrogen plasma). As-suming that $kT_2 \gg E_2$ (always valid for the case under investigation) we can approximate $\delta$ as

$$\delta = \frac{E_2 - E_1}{kT_2} \qquad (4)$$

$$= \frac{16(E_2 - E_1)}{5k\mathcal{M}^2 T} \qquad (5)$$

In terms of the initial (unshocked) ICM parameters

$$L = \frac{64}{k\sqrt{5}} \mathcal{L} V \mathcal{M}^{-1} n^2 T^{-1/2} (E_2 - E_1) \qquad (6)$$

Evaluating this expression over the *ROSAT* bandpass with the parameters relevant to Cygnus-A gives a luminosity of $7 \times 10^{39}$ erg s$^{-1}$. This is two orders of magnitude smaller than the observed emission. Thus, the observed X-ray emission cannot originate from smooth ICM that has been shocked by the radio lobe. Note that the presence of metals in the ICM would be expected to increase the emission slightly above this value.

Inhomogeneities in the ICM could significantly enhance the observed emission. Dense clouds (condensed from the cooling flow) might dominate the X-ray emission once the ICM has been compressed and heated by the shock. Consider a simplified two-phase model for the ICM consisting of cold dense clouds in pressure equilibrium with a hot connected phase. Suppose the clouds have a volume filling factor $f$ and (number) density given by $n_c = an$ (i.e. $a$ is the density contrast between the clouds and the hot phase of the ICM). Pressure equilibrium then gives the temperature to be $T_c = T/a$. The shock will propagate in the hot phase and envelope the clouds. The external pressure on the clouds will jump from $p$ to $p_2$ where

$$p_2 = \frac{5}{4} \mathcal{M}^2 p \qquad (7)$$

A shock is driven into the cloud with a Mach number (with respect to the unshocked *cloud* material) close to $\mathcal{M}$. Since the thermal emission observed in the *ROSAT* band varies as $n^2 T^{-1/2}$, the ratio of the shocked cloud emission to the emission from the shocked smooth ICM is

$$\mathcal{R} = f a^{5/2} \qquad (8)$$

We require $\mathcal{R} \sim 100$ in order to explain the observed emission. Models for a cooling flow lead to the relation $f = a^{-3+\alpha}$ (Thomas, Fabian & Nulsen 1987) where $\alpha$ is related to the cooling function and lies in the range $-0.5 < \alpha < 0.5$. Note that we have used an integrated form of the volume filling fraction given in Thomas, Fabian & Nulsen (1987). For thermal bremsstrahlung, $\alpha = 0.5$. This gives

$$\mathcal{R} = a^{-1/2+\alpha} \sim 1 \qquad (9)$$

Thus, it is not possible to have the required amount of mass in the form of cold dense clouds in order to explain the observed emission as arising from a shocked inhomogeneous ICM.

## 6 CONCLUSIONS

Thermal emission from the ICM dominates the observed X-ray flux in the *ROSAT* band. This emission is strongly peaked on Cygnus-A itself and extends over 0.5 Mpc in all directions from Cygnus-A. To the north-west is a 'plume' of material extending over 1.5 Mpc. The existence of this plume shows that the outer regions of the cluster are not dynamically relaxed. Indeed, this may be evidence for a recent (or on-going) cluster-cluster interaction or merger. Within the cluster merger scenario, the fact that the plume appears hotter than much of the rest of the cluster is puzzling given the well known positive correlation between cluster luminosities and cluster temperatures. Further studies of the thermal structure of this plume with instruments such as *ASCA*, together with a careful examination of the galaxy distribution within this cluster, may help resolve the origin of this plume.

Performing a deprojection analysis of the PSPC and HRI data confirms the existence of a significant cooling flow with an integrated mass deposition rate of $\sim 250$ M$_\odot$ yr$^{-1}$ and a (Hubble time) cooling radius of $\sim 180$ kpc. The central ICM density derived from the deprojection is $\sim 0.03$ cm$^{-3}$. The HRI data well resolve the spatial region over which the mass deposition occurs. A large fraction of the mass is deposited in the inner 50 kpc of the cluster, significantly less than the Hubble time cooling radius. This is interpreted as evidence for a disruption of the cooling flow approximately $3 \times 10^9$ yr ago. Further confirmation for the cooling flow comes from fitting thermal plasma models to spectra extracted in annuli centred on Cygnus-A. Such fits show a clear temperature gradient with the temperature decreasing from $> 4$ keV in the 4–5 arcmin radius annulus to $2.5^{+0.7}_{-0.4}$ keV in the central 1 arcmin radius. Unfortunately, the severe Galactic absorption to this source limits the constraints we can place on the plasma temperatures and leaves the plasma abundance completely unconstrained.

Within the central arcminute or so we see features due to the presence of the radio source. Flattening of the innermost X-ray isophotes at a position angle *perpendicular* to the radio axis is readily interpreted in terms of cavities in the ICM due to the radio lobes (Carilli, Perley & Harris 1994) whereas the elongation of isophotes slightly further from the centre is likely due to X-ray emission from the hot spots (HCP). We find the X-ray emission from the western hot spot to be harder than predicted by the SSC model of HCP. However, changing the geometry and/or size of the radio hot spot assumed in the SSC model can lead to a shift in the spectral break of the inverse Compton scattered flux and may reconcile the model with the current spectral data. Alternatively, if the hot spots harbour a population of low energy electrons extra to those inferred from current radio observations, inverse Compton scattering of relativistically beamed optical quasar emission may produce the observed X-ray emission. Other models, in particular thermal models, are rejected on physical grounds.

In the near future, other instrument promise to provide a wealth of detail about the various X-ray emissions from Cygnus-A. *ASCA* data will provide the first detailed information on the thermal structure and abundances of the ICM. The hard X-ray mission *XTE* will be able to clearly detect and probe details of the primary AGN emission. Then, near the end of the decade, *AXAF* will provide unprecedented spatial resolution (together with spectral data). Such data will be invaluable in unambiguously disentangling such a complex system.

**ACKNOWLEDGMENTS**

We wish to thank Steve Allen, Niel Brandt, Carolin Crawford, Harald Ebeling, Alastair Edge, Roderick Johnstone and Dave White for many useful conversations and much help throughout the course of this work. The *ROSAT* HRI image used in our deprojection was kindly supplied by Dan Harris. CSR thanks PPARC for support and ACF thanks the Royal Society for support.